\documentclass[showpacs,prl,twocolumn]{revtex4}

\usepackage{epsfig}
\usepackage{amsmath}
\usepackage{graphicx}
\usepackage{dcolumn}
\usepackage{amsmath}
\usepackage{latexsym}
\usepackage{color}
\begin{document}

\title{Ulta-slow relaxation in discontinuous-film based electron glasses}

\author{T. Havdala}
\author{A. Eisenbach}
\author{A. Frydman}
\address{The Department of Physics, Bar Ilan University, Ramat Gan 52900,
Israel}

\begin{abstract}

We present field effect measurements on discontinuous 2D thin films
which are composed of a sub monolayer of nano-grains of Au, Ni, Ag or Al. Like other
electron glasses these systems exhibit slow conductance relaxation
and memory effects. However, unlike other systems, the discontinuous
films exhibit a dramatic slowing down of the dynamics below a
characteristic temperature $T^*$. $T^*$ is typically between 10-50K and is sample dependent.
For $T<T^*$ the sample exhibits a few other
peculiar features such as repeatable conductance fluctuations in
millimeter size samples. We suggest that the enhanced system sluggishness is related to the
current carrying network becoming very dilute in discontinuous films so that the system contains
many parts which are electrically very weakly connected and the transport is dominated by very few weak links. This enables studying the glassy properties of the sample as it transitions from a macroscopic sample to a mesocopic sample, hence, the results provide new insight on
the underlying physics of electron glasses.
\end{abstract}


\date{\today}

\maketitle

Glassy behavior of the conductivity, $\sigma$, in strongly
disordered systems that are characterized by strong electronic
interactions were predicted by several groups \cite{Grunewald,
pollak1, pollak2, davies, vignale}. Exciting such a system out of
equilibrium leads to an increase in conductivity, $\sigma$, after
which the relaxation towards equilibrium is characterized by
extremely long times, memory phenomena and aging. Since the slow
dynamics are related to their electronic properties these systems
were termed electron glasses \cite{davies}. Experimentally, glassy
features were observed in a verity of systems including granular Au
\cite{adkins}, amorphous and poly-crystalline indium oxide films
\cite{moshe,ady0,ady1,ady1_5,ady2}, ultrathin Pb films
\cite{goldman}, granular aluminum \cite{grenet1, grenet2} and thin
beryllium films \cite{be}. A standard way of excitation in these
experiments is by  applying a gate voltage, $V_{g}$, in a MOSFET
setup. Conductivity increases for both orientations of $V_{g}$
followed by very slow relaxation of $\sigma$ which is found to
follow an approximate logarithmic dependence on time and may be
measured over time-scales of days. A typical feature which has been
suggested as the hallmark of \emph{intrinsic} electron glasses
\cite{zvi_dip} is a "memory dip" (MD) which shows up as a minimum in the
$\sigma(V_g)$  curve when $V_{g}$ is scanned fast compared to the
characteristic relaxation time.  The dip is centered around the gate
voltage at which the sample was allowed to equilibrate.

The origin of the extremely slow relaxation and the memory dip as
well as their dependence on parameters such as temperature, bias
voltage, carrier concentration etc. are still under debate and more
experimental information may help shedding light on the physics of
electron glasses. In this letter we present results on the glassy
properties of two dimensional discontinuous films. We find that these systems
exhibit a dramatic slowing down of the dynamics below a
characteristic temperature $T^*$. For $T<T^*$ the conductance of the
sample exhibits reproducible fluctuations with exponentially growing
amplitude as the temperature is lowered indicating that the effective electronic size of the sample has become very small. We discuss the influence of
the sample geometry and its effective size on the conductance relaxation properties.

All conductance results presented in this work were obtained by standard 2-wire lock-in techniques performed on thin discontinuous
Au, Al, Ag or Ni films prepared by the quench condensation technique, i.e. thermal evaporation on a cryocooled substrate
\cite{strongin,aviad1,aviad2}. dc techniques were used for comparison in a few cases yielding similar results. For
achieving field effect geometry we used a doped Si substrate (that
was utilized as a gate electrode) coated by a 0.5 $\mu m$ insulating
SiO layer. Gold pads were pre-prepared on the substrate so that,
together with a shadow mask, they defined a sample area of 0.6mm by
0.6mm. The substrate was then placed on a sample holder within a
vacuum chamber. After the chamber was pumped out, the substrate was
cooled to cryogenic temperatures and thin layers of Au, Ni, Al or Ag were
deposited while monitoring the film thickness and resistance. For
thin enough layers this technique yields a film that is
discontinuous, consisting of a sub monolayer of metallic grains, 10-20nm in
diameter, separated by vacuum as seen in fig. 1. A major advantage of this method is that throughout the entire process of sample growth and
measurement the samples are kept in ultra-high vacuum and not
exposed to air, thus protecting the grains from oxidation or
contamination. This is especially important for nano-grains in which
the surface area to volume ratio is very high.

\begin{figure}[t]
\vspace{2cm}
    \centering
    \includegraphics[width=\columnwidth,keepaspectratio=true]{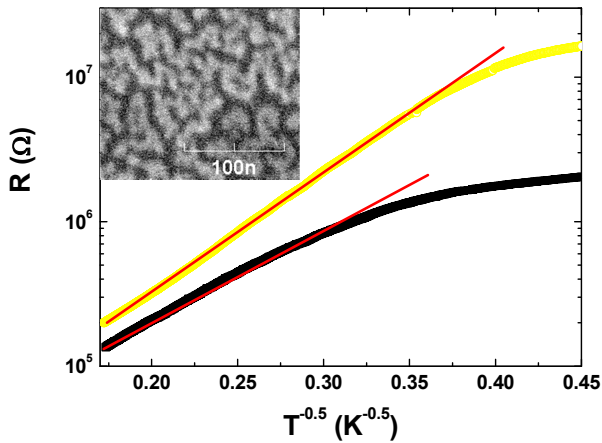}
    \vspace{-2.5cm}
    \caption {Resistance versus temperature for a 7 nm thick discontinuous gold film (black full squares)measured using dc methods and a 10nm Ni Film (yellow empty circles) measured with ac methods. The lines are fits to Efros Shklovskii like behavior. Insert: An SEM image of the Au film after heating it to room temperature (color online).
    }
    \vspace{-0.5cm}
\end{figure}

Granular metals which are on the insulating side of the
metal-insulator-transition are known to be hopping systems that exhibit
resistance versus temperature (R(T)) curves that follow:
\vspace{-0.2cm}
\begin{equation}
\label{eq_RT} R\propto exp[\frac{T_{0}}{T}]^{\alpha}
\end{equation}

Experimentally, $\alpha s$ between 0.5 and 1 have been reported. In
our films eq. \ref{eq_RT} is fulfilled only for relatively high
T. Fig. 1. depicts the R(T) curves of a gold film and a nickel film. It is seen
that for temperatures higher than a characteristic temperature,
$T^*$ (in fig 1 $T^* \simeq 20K$ for the gold sample and $\simeq 10K$ for the Ni sample), we observe a usual Efros
Shklovskii \cite{ES} like dependence ($\alpha=0.5$). However, for
$T<T*$ the conductance depends much weaker on temperature and seems to be approaching saturation at low temperatures.

It turns out that T* manifests itself in other transport properties
of the sample. As in other electron glasses, after the sample is
allowed to equilibrate at a certain gate voltage, $V_{g0}$  for
a long time, a MD  is observed in conductance versus gate voltage
curve. In all our films the MD is accompanied by reproducible conductance
fluctuations (see fig. 2). These fluctuations are random, however they are reproducible for sequential experiments performed under similar conditions. Such fluctuations have been observed in the past in indium oxide \cite{vladimeso} and granular Al
\cite{grenetmeso} electron glasses with dimensions smaller than $100
\mu m$. We observe these fluctuations for much larger samples. Fig.
2 shows that the rms amplitude of the fluctuations, $\frac{\delta G}{<G>}$,
of a 0.6*0.6 mm film becomes measurable for $T<T^*$ and grows
exponentially with decreasing temperature. The increase of $\frac{\delta G}{<G>}$
 with lowering T was observed in other systems \cite{grenetmeso}, however, in those cases it was found that
 $\frac{\delta G}{<G>} \propto T^{ - \beta} $ with $\beta \sim 2$. The power law
dependence was attributed to the fact that the magnitude of the
fluctuations are proportional to $L_{0}^{2}$ where $L_{0}$ is the
spatial scale of an independent microscopic fluctuations which was ascribed
to the correlation length of the percolation network. This length
is predicted to have a power law dependence on temperature
\cite{hill,pollak,shklovskii}

\vspace{-0.5cm}

\begin{equation}
\label{eq_L0} L_0 \propto [r_h]^\nu \propto T^{-\alpha \nu}
\end{equation}

\vspace{-0.2cm}

where $r_h$ is the characteristic hopping length, $\alpha$ is the
power of the temperature dependence of the resistance given by eq.
\ref{eq_RT} and $\nu$ is between 1 and 2. An exponential dependence
of $\frac{\delta G}{<G>}$ as seen in fig. 2 implies an unconventional change of the percolation network for $T<T^*$.

\begin{figure}
\vspace{-1cm}
    \centering
    \includegraphics[width=\columnwidth,keepaspectratio=true]{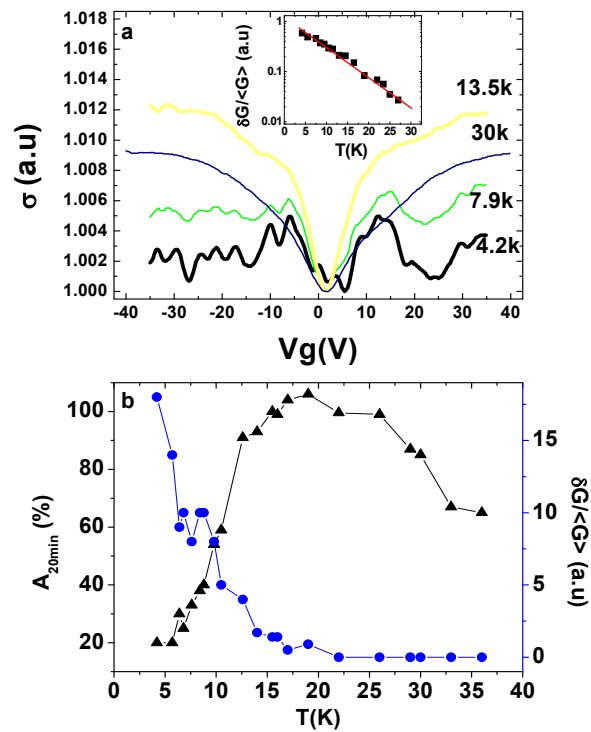}
    \vspace{-1cm}
    \caption {a: Conductance as a function of gate voltage for one of our
    0.6*0.6 mm Au films at different temperatures. The thickness was 7nm. $V_{g0}=0V$. The inset shows the rms amplitude of the conductance fluctuations as a function of temperature.
    b: $\frac{\delta G}{<G>}$ (blue circles) and the dip amplitude 20 minutes after the quench-coolling of the sample (black triangles) versus T.
    (color online)
    }
    \vspace{-0.5cm}
\end{figure}

A similar conclusion can be derived from the temperature dependence
of the current-voltage characteristics. It is useful to study
conductivity versus voltage curves, $\sigma(V)$, such as that shown
in fig. 3  for an Au film. These curves are characterized by a
voltage $V_{0}$ above which the conductivity increases with voltage
thus deviating from ohmic behavior. Several theoretical works
\cite{hill,pollak,shklovskii} predict that non-Ohmic conductivity
should occur for:
\vspace{-0.6cm}

\begin{equation}
\label{nonohmic} k_{B}T<eFL_{0}
\end{equation}

where F is the field applied across $L_0$. Hence $V_0$ is expected
to follow \cite{iv}:
\vspace{-0.3cm}
\begin{equation}
\label{V0} V_0 \propto \frac{T}{L_0} \propto T^m
\end{equation}
where m is between 1 and 2. The dependence of $V_0$ on T for a discontinuous Au film is shown in
fig. 3. It is seen that for $T>T^*$ the I-V characteristic yields $V_0 \propto T^2$ as expected.
For $T<T^*$ there is a sharp drop in $V_0$ and a clear deviation
from eq. \ref{V0}. Such behavior could be interpreted as an unusually  rapid
increase in $L_{0}$ as the temperature is lowered below $T^*$.

\begin{figure}
\vspace{1.7cm}
    \centering
    \includegraphics[width=\columnwidth,keepaspectratio=true]{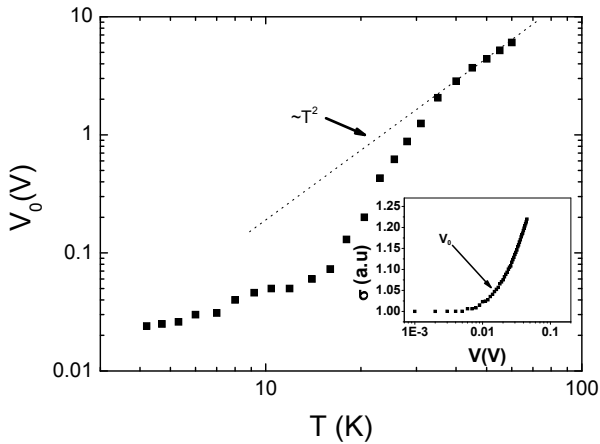}
    \vspace{-2.9cm}
    \caption {$V_0$ as a function of T for a discontinuous Au film. The determination of $V_0$ is illustrated in the inset which shows a conductance versus voltage
    curve at T=6K. $V_0$ is defined as the voltage at which the conductance increases by 5 \% from its ohmic value.}
    \vspace{-0.6cm}
\end{figure}

Perhaps the most remarkable phenomenon that occurs at low temperatures
is a dramatic slowing down of the system's electronic dynamics.
Since the relaxation after any excitation follows an approximate
logarithmic dependence it does not have a natural characteristic
decay time. Nevertheless a number of methods have been proposed to
experimentally define the "slowness" of the relaxation in electron
glasses \cite{zvi_tau}, all giving equivalent characteristic times,
$\tau$. A popular method employs the "two dip experiment" (TDE)
\cite{ady2}. In this experiment the sample is allowed to equilibrate
for a long time (of the order of a day) at a certain gate voltage
$V_{g1}$. At this stage a fast $\sigma(Vg)$ scan yields a
conductivity memory dip centered around $V_{g1}$. At time t=0 the
gate voltage is abruptly changed to $V_{g2}$ and fast $\sigma(V_g)$
scans are performed at selected time intervals. As a function of
time the dip at $V_{g1}$ is slowly suppressed while a new dip
develops around $V_{g2}$. The characteristic relaxation time,
$\tau$, is defined as the time at which the amplitude of the two
dips is equal.

A TDE for a discontinuous Au film sample at T=55K is shown in fig. 4e.  It is seen
that $\tau _{55k} \approx 1000 s$, which is a typical value for
disordered samples having high carrier concentration \cite{zvi_dip}. Upon lowering the temperature, all our films exhibits a huge
increase in the characteristic relaxation time. For example, $\tau _{15k}$ is
found to be $\sim 10^6 s$. This makes the systematic study of
$\tau_{TDE}$ at different temperatures unpractical. Therefore we
measure the fraction of relaxation that the system undergoes over a
time of 1 hour. This is done in the following way: We allow the
sample to equilibrate at relatively high T ($T>T^*$) while applying
$V_{g}$=0 for a time $t_{1}$ after which a dip is well developed. We then
cool the sample to a different temperature, perform a gate voltage
sweep to determine the size of the dip and change the gate voltage
to $V_{g}=-13V$. We then wait for one hour and measure the amplitude
of the dip at $V_{g}=-13$. We define the amplitude ratio between the new dip and the old dip (at $V_{g}=0$) as $A_{1h}$. This is taken as a measure for the "slowness" of the relaxation. The dependance of $A_{1h}$ on temperature is shown in fig. 4. It is seen
that there is a dramatic decrease of $A_{1h}$ at $T \sim T^*$. Similar behavior was obtained for all studied samples (over 12). The relaxation times and $T^*$ did not seem to depend on resistance but the amplitude of the memory dip decreased as the sample approached the metal-insulator transition.

\begin{figure}
\vspace{-0.9cm}
    \centering
    \includegraphics[width=\columnwidth,keepaspectratio=true]{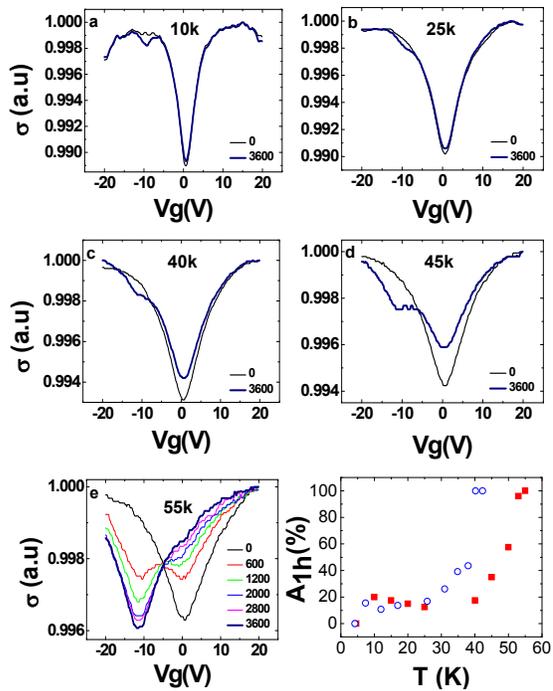}
    \vspace{-1.3cm}
    \caption {a-d: Conductance versus gate voltage at t=0 (black light
    line) and at t=1 hour (blue heavy line) for different
    temperatures. e: More detailed two dip experiment results for T=55K. f: Second dip amplitude of the Au sample of panels a-e (red full squares) and a Ni sample (blue empty circles)as a function of temperature (color online).
    }
    \vspace{-0.3cm}
\end{figure}

Slowing down of relaxation processes with decreasing temperature may seem natural, however it is
contrary to the situation in other electron glass systems. In
amorphous and crystalline indium oxide \cite{zvi_T}, and granular
aluminum \cite{grenet2} the dynamics were found either to be
independent on temperature or to slow down upon increasing T. The
latter has been suggested as evidence that disordered electronic
systems are quantum glasses \cite{zvi_T}. In contrast, our Au films
exhibit a dramatic \emph{slowing} of the dynamics upon cooling.
Furthermore, the temperature dependence of $\tau$ is very peculiar.
Fig. 4f depicts $A_{1h}$ (which inversely depends on $\tau$) as a function of T.  It is seen that there is
a sharp increase of relaxation times over a small temperature range.

The fact that the slowing down of the dynamics occurs at $T \sim T^*$,
where mesoscopic conductance fluctuations become significant, leads us to postulate that it
is related to a significant dilution of the current carrying network (CCN). Note that, unlike in other electron glasses,
the conductivity in the discontinuous films is strictly two dimensional. In addition, the SEM image of figure 1 shows that the grains are not closely packed, but rather the film morphology is composed of fractal shaped clusters connected by thin (single grain) bottlenecks \cite{rem}. Hence this film is characterized by geometrical and not only electronic percolation which is typical of other hopping systems. At low temperatures many of these bottlenecks may disconnect from the CCN because of energy mismatches thus leaving most of the sample electrically cut-off. This gives rise to enhanced conductance fluctuations since the effective electronic system size is small. This also accounts for the apparent saturation of the R(T) curve at low T as shown in fig 1. Upon lowering T the electric current network becomes progressively dilute and the variable range hopping mechanism becomes less relevant as fewer sections of the sample dominate the conductivity. Eventually, at the extreme case where the conductivity governed by a single weak link it is expected to be temperature independent.

Under these conditions, at low T there is a very small probability for an electron to tunnel into and out of the current carrying
network and most of the charge carriers are trapped in isolated regions of the sample. This causes a significant slowing down of the relaxation processes to equilibrium because these rely on many body hopping processes taking place in various parts of the sample. Since most of the sample is very weakly connected, the relaxation to equilibrium becomes extremely slow, thus hindering the development of a new MD at low temperatures. This can be realized from fig. 2b that shows the amplitude of the memory dip 20 minutes after the cool down, $A_{20m}$, as a function of temperature. It is seen that for $T<T^*$ this amplitude reduces with decreasing temperature. The amplitude of the  memory dip in other electron glasses has shown to increase rapidly with decreasing temperature \cite{ady0,grenet2}. The decrease of $A_{20m}$ with lowering temperature reflects the fact that the dynamics of the disconnected sections (most of the sample) have become extremely slow that the CCN is so dilute so that glassy properties are no longer relevant. Indeed some of the mesoscopic electron glasses studied in the past also did not exhibit a measurable MD \cite{vladimeso, grenetmeso}.

In conclusion we have shown that electron glasses based on 2D discontinuous films exhibit a sharp increase of relaxation times at
low temperatures. These systems are characterized by a tenuous morphology which was not studied in the context of electron glasses so far. Our results demonstrate that relaxation to equilibrium hinges upon electronic transition in a wide region of the sample and when these become unavailable the relaxation processes are considerably impeded. The many-body and many-electron-hopping nature of the electron glass thus becomes strikingly apparent in these discontinuous films.

We are grateful for useful discussions with A. Amir, Z. Ovadyahu and
M. Pollak. This research was supported by the Israeli academy of
science (grant number 399/09)


\begin{references}

\bibitem{Grunewald} M. Grunewald, B. Pohlman, L. Schweitzer, and D. Wurtz, J.
Phys. C \textbf{15}, L1153 (1982).
\bibitem{pollak1} M. Pollak and M. Ortuno, Sol. Energy Mater. \textbf{8}, 81 (1982);
\bibitem{pollak2} M. Pollak, Philos. Mag. B \textbf{50}, 265 (1984).
\bibitem{davies} J. H. Davies, P. A. Lee and T. M. Rice, Phys. Rev. Lett. \textbf{49}, 758
(1982).
\bibitem{vignale} G. Vignale, Phys. Rev. B 36, 8192 (1987).

\bibitem{adkins}  C.J. Adkins, J.D. Benjamin, J.M.D. Thomas, J.W. Gardner, A.J.Mc Geown, . J. Phys. C  \textbf{17}, 4633 (1984).
\bibitem{moshe} M. Ben-Chorin, D. Kowal and Z. Ovadyahu, Phys. Rev. \textbf{B44}, 3420 (1991).
\bibitem{ady0} A. Vaknin,  Z. Ovadyahu and M. Pollak, Europhys. Lett.,
\textbf{42} 307 (1998)
\bibitem{ady1}A. Vaknin,  Z. Ovadyahu and M. Pollak, Phys. Rev. Lett. \textbf{81}, 669 (1998).
\bibitem{ady1_5} A. Vaknin, Z. Ovadyahu and M. Pollak, Phys. Rev.
Lett. \textbf{84}, 3402 (2000).
\bibitem{ady2}A. Vaknin,  Z. Ovadyahu and M. Pollak, Phys. Rev. \textbf{B65}, 134208 (2002).
\bibitem{goldman} G. Martinez-Arizala, D.E. Grupp, C. Christiansen, A.M.
Mack, N. Markovic, Y. Seguchi, A.M. Goldman, Phys.
Rev. Lett. \textbf{78}, 1130 (1997)
\bibitem{grenet1} T. Grenet, Eur. Phys. J. B \textbf{32}, 275 (2003).
\bibitem{grenet2} Grenet et al., Eur. Phys. J. B \textbf{56}, 183 (2007).
\bibitem{be} Z. Ovadyahu, Y.M. Xiong and P.W. Adams, Phys. Rev. B., \textbf{82}, 195404, (2010)

\bibitem{zvi_dip} Z. Ovadyahu, Phys. Rev. B \textbf{78}, 195120 (2008).
\bibitem{strongin} M. Strongin, R. Thompson, O. Kammerer and J. Crow, Phys.
Rev. {\bf B1}, 1078 (1970).

\bibitem{aviad1} A. Frydman and R.C. Dynes, Sol. State Comm. \textbf{110}, 485 (1999).

\bibitem{aviad2} A. Frydman, T.L. Kirk and R.C. Dynes, Solid State Commun. \textbf{114}, 481 (2000).

\bibitem {ES} A. L. Efros and B. I. Shklovskii, J. Phys. C \textbf{8}, L49 (1975); B. I. Shklovskii and A. L. Efros, Electronic Properties of
Doped Semiconductors (Springer, New York, 1984).

\bibitem{vladimeso} V. Orlyanchik and Z. Ovadyahu,Phys. Rev.\textbf{ B75}, 174205
(2007).

\bibitem{grenetmeso} J. Delahayea, T. Grenet, and F. Gay, Eur. Phys. J. \textbf{B65}, 5
(2008).
\bibitem{hill} R. M. Hill, Philos. Mag. \textbf{24}, 1307 (1971).
\bibitem{pollak} M. Pollak and I. Riess, J. Phys. C \textbf{9}, 2339 (1976).
\bibitem{shklovskii} B. I. Shklovskii, Sov. Phys. Semicond. \textbf{10}, 855 (1976)
\bibitem{iv} D. Talukdar, U. N. Nandi, K. K. Bardhan, C. C. Bof Bufon, T. Heinzel, A. De, and C. D.
Mukherjee, Phys. Rev. \textbf{B84}, 054205 (2011)
\bibitem{zvi_tau} Z. Ovadyahu, Phys. Rev. B \textbf{73},214208 (2006).
\bibitem{grenet_tau} T. Grenet and J. Delahaye, arXiv:1105.0984.
\bibitem{amir} A. Amir, Y. Oreg and Y. Imry, Annual Review of Condensed Matter, \textbf{2}, 235, (2011)
\bibitem{zvi_T} Z. Ovadyahu, Phys. Rev. Let. \textbf{99}, 226603 (2007).
\bibitem{rem} Though the morphology may change while heating the sample to room temperature we assume that the
qualitative structure is similar at low T.


\end{references}
\end{document}